\documentclass[11pt]{article}
\usepackage{jheppub}
\usepackage{tabularx} 
\usepackage{graphicx} 
\usepackage{mathtools}
\usepackage{amsmath}
\usepackage{amsfonts} 
\usepackage{listings}
\usepackage{bbold}

\title{On  quantum corrections to  BPS 
 Wilson loops in superstring theory on \boldmath $AdS_3\times S^3 \times T^4$ with mixed  flux}

\author{Daniel Pajer}

\affiliation{Blackett Laboratory, Imperial College London, London SW7 2AZ, U.K.}

\emailAdd{daniel.pajer17@imperial.ac.uk}

\abstract{We revisit the calculation of one-loop corrections to the partition functions of  a string ending on a circular or straight line in  type IIB string theory on $AdS_3\times S^3 \times T^4$  supported  by a mixture of  3-form fluxes depending on a parameter $q$ (with $q=0$ being R-R and $q=1$  pure NS-NS cases). Correcting earlier results in the literature, we find that the partition function  depends on $q$ only through rescaled tension $T= T_0/(1-q^2)$.}

\begin{document}
\maketitle

\section{Introduction}

One of the important pathways to probing the $AdS$/CFT conjecture has been through finding exact solutions on both sides of the duality by the means of integrability (see, e.g., \cite{Beisert_2011}). Arguably one of the most profound class of operators on the field theory side of the duality are the Wilson loops. These observables have a very natural representation on the string theory side of the duality; the Wilson loop is related to the minimal surface area swept out by a moving string whose two ends are tied to the conformal boundary \cite{Maldacena_1998, Aharony_2000}. 

If one wants to  calculate the Wilson loop 
using perturbative string theory beyond the leading (classical) order, one needs to introduce quantum fluctuations to the minimal surface solution. If we expand the Wilson loop  in terms of the string coupling, $g_s \to 0$, we find (for a review see  \cite{Giombi_2020})
\begin{equation}
\langle \mathcal{W} \rangle =  \frac{1}{g_s}Z + \mathcal{O}(g_s), \quad Z = \int \mathcal{D}[\text{fields}]\, e^{-S_{\text{GS}}},
\end{equation}
where $S_{\text{GS}}$ is the Green-Schwarz action and $Z$ is the  string partition function, defined as a path integral over all the fields.

We will consider quadratic perturbations in both the bosonic and fermionic fields around a  classical solution, which is equivalent to including the one-loop quantum interactions. If we write
\begin{equation}
\langle \mathcal{W} \rangle = e^{-\Gamma}, \quad \Gamma = \Gamma_0 + \Gamma_1 + \Gamma_2 + \cdots,
\end{equation}
then $\Gamma_0 \propto T_0$ is simply the value of the classical string action, $\Gamma_1 \propto (T_0)^0$ will be the one-loop quantum correction (which will include possible measure-related normalisation factors), and $\Gamma_2 + \cdots = \mathcal{O}(T_0^{-1})$ the higher order corrections which we will neglect in the low energy limit. The one-loop approximation to $\langle \mathcal{W} \rangle$ will be referred to as the semiclassical partition function and evaluating it will be the aim of this work. 

Here we shall study the string theory on the $AdS_3\times S^3\times T^{4}$ background  supported by a combination of R-R and NS-NS 3-form fluxes  parametrised by $q$ (see, e.g., \cite{Cagnazzo_2012,Hoare:2013pma} and references there). We will focus on the two simple cases of a straight-line and circular loop. These have been already discussed in \cite{Drukker_2000} in the absence of the NS-NS flux ($q=0$). For  general  $q$ this problem   was addressed  earlier in \cite{Hernandez_2020}, but we believe  the discussion in that paper contained several flaws that  will be corrected below.

\section{Classical solutions}
In this section, we will consider a type IIB superstring on the $AdS_3\times S^3\times T^{4}$ background supported by a combination of 3-form R-R and NS-NS fluxes, described later in (\ref{fluxes}), and focus on finding the associated classical solutions. Yet for the discussion of these solutions, one only needs to consider the presence of the metric and of the NS-NS $B$-field flux. Furthermore, we shall focus solely on solutions located in the $AdS_3$ subspace, and thus only the $AdS_3$ part of the NS-NS field will be needed in this section. In the case of the $AdS_5\times S^5$ background (without the NS-NS flux), a general class of classical solutions representing minimal surfaces ending at the boundary  was found in \cite{Drukker_2006} using periodic ansatze which helped reduce the complex dynamical system to a one-dimensional integrable system. A similar analysis  was later performed in the case of the $AdS_3\times S^3\times T^{4}$ background supported also with the flux of  NS-NS  $B$-field  \cite{Hernandez_2019}. We shall  follow this reference to find the two simple cases of a string ending on a circular loop or a straight line at the boundary of the $AdS_3$ space. These simple solutions  were   not properly identified in a subsequent  work \cite{Hernandez_2020}  and our  aim in this section will be to present them in a clear way.

\subsection{Single circle}
The solution  we shall discuss will be located in the (Euclidean) $AdS_3$ part of $AdS_3\times S^3\times T^{4}$
so the rest of the coordinates can be ignored.
To describe the $AdS_3$ part of the metric, we will use the standard Poincaré coordinates, however in their polar form, i.e.
\begin{equation}
\mathrm{d}s^2_{AdS} = \frac{\mathcal{R}^2_{AdS}}{z^2} \big(r^2 \mathrm{d}\phi^2 + \mathrm{d}r^2 + \mathrm{d}z^2\big)
\end{equation}
with $\phi \in [0,2\pi)$, $r \in [0,\infty)$, $z \in (0,\infty)$, hence our coordinates are now dimensionless, with $\phi$ being the Euclidean time-like coordinate, and $\mathcal{R}_{AdS}$ is the $AdS$ radius.

The bosonic string action will read
\begin{equation}
\label{circleaction}
\begin{split}
S &= \frac{T_0}{2} \int  \mathrm{d}^2\sigma \Big(\delta^{ij} G_{\mu\nu} \partial_i X^\mu \partial_j X^\nu + \epsilon^{ij} B_{\mu\nu} \partial_i X^\mu \partial_j X^\nu \Big),
\end{split}
\end{equation}
where we have set the conformal gauge, $\sqrt{-\gamma}\gamma^{ij} = \eta^{ij}$, and subsequently Wick rotated the time-like string coordinate such that $\eta^{ij} \to \delta^{ij}$, furthermore the antisymmetric field\footnote{Apart from the standard first term of the field, we have also added the second term to provide a natural regularisation of the action. This term is a total derivative and thus does not influence the equations of motion. A different choice of gauge can result in different value of the action, yet this choice also seems to be consistent while using the coordinates $\{u,v\}$ defined in (\ref{eqcoordinates}), where the field would naturally become $B \propto \mathrm{d}v \wedge \mathrm{d}\phi$, with $H = \mathrm{d}B$ unchanged, resulting in the same equations of motion and value of the action (see appendix \ref{appendix1}).} is
\begin{equation}\label{bfield}
B = \frac{q \mathcal{R}^2_{AdS}}{2z^2}\,\Big[\mathrm{d}(r)^2 + \mathrm{d}  (z-qR)^2  \Big] \wedge \mathrm{d}\phi,
\end{equation}
where we have define the dimensionless constant $R$ such that
\begin{equation}
R  \equiv \frac{1}{\sqrt{1-q^2}},
\end{equation}
with $q\in (-1,1)$\footnote{In the case of pure NS-NS flux (i.e. when $|q|=1$), our solution would become singular. The corresponding string theory in this  case  is  described by the WZW model (see, e.g., \cite{Maldacena:2000hw}). The classical action trivialises, as the surface becomes flat, and so should the partition function, as all the bosonic and fermionic excitations become massless in this limit. In this case  the worldsheet adheres to the boundary of the $AdS_3$ space.}. This choice of the $B$-field is consistent as the 3-form NS-NS field takes the form
\begin{equation}
H = \mathrm{d}B = 2q\mathcal{R}_{AdS}^2 \frac{r}{z^3}\, \mathrm{d}\phi \wedge \mathrm{d} r \wedge \mathrm{d} z.
\end{equation}

Now to simplify the expressions we will redefine $T_0$ such that it absorbs the $\mathcal{R}^2_{AdS}$ factor, i.e. take $T_0 \to  \mathcal{R}^2_{AdS} T_0$, $G_{\mu\nu} \to \mathcal{R}^{-2}_{AdS} G_{\mu\nu}$, and $B_{\mu\nu} \to \mathcal{R}^{-2}_{AdS} B_{\mu\nu}$. Thus $T_0$ is from now on a dimensionless parameter in natural units.

Following \cite{Drukker_2006,Hernandez_2019}, we will use the periodic ansatz which is consistent with the equations of motion,
\begin{equation} \label{eqBC3} 
\phi(\tau,\sigma) = \tau, \quad
r(\tau,\sigma) = r(\sigma), \quad  z(\tau,\sigma) = z(\sigma),
\end{equation}
alongside with the boundary conditions defining the unit-radius circular Wilson loop
\begin{equation} \label{eqBC1} 
r(\tau,\sigma_i) = 1, \quad  z(\tau,\sigma_i) = 0, \quad 
i = 1,2, 
\end{equation}
with $\tau \in [0,2\pi)$ and $\sigma \in [\sigma_1,\sigma_2]$.

Using this ansatz, we find the action to take the form
\begin{equation}
\begin{split}
S &= \frac{T_0}{2} \int  \mathrm{d}^2\sigma\, \frac{1}{z^2}\Big(r'^2 + z'^2 + r^2 - 2qrr' - 2qzz' + 2q^2 R z'\Big),
\end{split}
\end{equation}
where $r'\equiv \mathrm{d}r/\mathrm{d}\sigma$ and $z'\equiv \mathrm{d}z/\mathrm{d}\sigma$. The integrability of this model follows from the existence of the first integral of motion 
\begin{equation}\label{eqC3}
p = z^{-2} (rr' + zz' - qr^2),
\end{equation}
supplied with the conformal gauge condition requiring 
\begin{equation} \label{eqC2} 
r'^2 + z'^2 - r^2 = 0.
\end{equation}
These two equations together with the boundary conditions (\ref{eqBC1}) reduce our problem to the desired 1d integrable model. Furthermore, we find the circular solution by setting $p=0$, which seems to be a special case which would otherwise for $p\neq 0$ describe a minimal surface ending on two concentric circles.\footnote{In \cite{Hernandez_2019} one can find a very brief discussion involving the single circular case (i.e. when $p=0$), yet this case is argued to be the single circle solution only in the pure R-R flux setting (when $q=0$ as in e.g. \cite{Drukker_2006}). The authors argue that even if $p=0$ but $q\neq 0$, the resulting surface is that of one ending on two concentric circles, which cannot fit inside an $AdS_2$ subspace. Instead we argue that the $p=0$ case is not a special limit of $p\neq 0$ but rather a different case entirely, and that it results in an $AdS_2$ surface ending on a single circle for any value of $q$.}

To solve this model, we introduce a new coordinate system, $\{r,z\}\to\{u,v\}$, which proves to be easier to work with than the one used in \cite{Hernandez_2019}
\begin{equation} \label{eqcoordinates} 
\begin{split}
r &= \frac{e^v}{\sqrt{1+u^2}} ,\\
z &= \frac{ue^v}{\sqrt{1+u^2}} + qR,\\
\end{split}
\end{equation}
In these coordinates, we find that (\ref{eqC3}) and (\ref{eqC2}) become, respectively,
\begin{equation}
\begin{split}
v'(1+u'^2) + qR e^{-v} \sqrt{1+u^2}\bigg(uv' +\frac{u'}{1+u^2}\bigg) - q &= 0,\\
v'^2(1+u'^2) + \frac{u'^2}{1+u^2} &= 1.
\end{split}
\end{equation}
After rearranging, one finds this set of equations reduces to
\begin{equation}
\begin{split}
v' &= 0,\\
u'^2 &= 1 + u^2.
\end{split}
\end{equation}
These equations, alongside with the boundary conditions, are satisfied by
\begin{equation}
\begin{split}
e^v &= R,\\
u(\sigma) &= \mathrm{sinh}\big( \sigma - \mathrm{arctanh}\,q \big),
\end{split}
\end{equation}
with $\sigma \in [0,\infty)$. This in turn lets us express $r$ and $z$ in a more elegant form, using $R = 1/\sqrt{1-q^2}$, as
\begin{equation}
\label{rz}
\begin{split}
r &= R\,\,\mathrm{sech} \big(\sigma - \mathrm{arctanh}\,q\big),\\
z &= R\,\Big[\mathrm{tanh} \big(\sigma - \mathrm{arctanh}\,q\big) + q \Big].
\end{split}
\end{equation}

Looking at (\ref{rz}), we see that $r$ and $z$ can be related by a very simple, yet powerful expression, 
\begin{equation}
\label{circle}
r^2 + (z-qR)^2 = R^2, 
\end{equation}
which is just the equation for a circle of radius $R$ whose centre is shifted in the $z$-direction. If one were to compare this to the case with vanishing $B$-field, i.e. $q=0$, found in \cite{Drukker_2006}, one would see that the $B$-field acts to push the centre of the spherical minimal surface up or down, depending on the sign of $q$, as well as increasing its radius, from 1 to $R$. The geometry of this minimal surface is thus again that of $AdS_2$, which can be seen easily when the target metric is pulled back onto the worldsheet as
\begin{equation}
\mathrm{d}s^2 = R^2 \, \frac{\mathrm{d}\tau^2 + \mathrm{d}\sigma^2}{\mathrm{sinh}^2\,\sigma},
\end{equation}
and the presence of the $q$ term seems to only increase the radius of the induced metric by factor $R$.

Additionally, we are able to explicitly evaluate the action, to this end, we shall use (\ref{eqC2}) and (\ref{circle}) to simplify the Lagrangian which leaves us with the standard kinetic term, 
\begin{equation}
\begin{split}
S = T_0 \int \mathrm{d}^2\sigma \, \frac{r^2}{z^2}. 
\end{split}
\end{equation}
Utilising the fact that $r/z = R/\mathrm{sinh}\,\sigma$, we find this integral reduces to
\begin{equation}
\begin{split}
S = T_0 R^2 \int \mathrm{d}^2\sigma \, \frac{1}{\mathrm{sinh}^2 \sigma} = 2\pi T_0 R^2 (\mathrm{coth}\,\epsilon - 1), 
\end{split}
\end{equation}
where we have introduced the cut-off, $\epsilon \to 0$, to regularise the action as the integral is divergent (see, e.g., \cite{Semenoff_2002}). Finally, the regularised action for the circular case is given by 
\begin{equation}
\Gamma_0 = -2\pi T_0 R^2 = -\frac{2\pi T_0}{1-q^2}.
\end{equation}
Interestingly, one finds that the same (regularised) action could have been found by setting $B = 0$ and rescaling the $AdS$ metric by $R^2$, i.e. $\mathrm{d}s^2_{AdS} \to R^2 \mathrm{d}s^2_{AdS}$.

\subsection{Straight line}
We will now focus on the case when the string ends on a straight line on the conformal boundary. This case is quite straight-forward, as it carries many similarities with the circular case but is algebraically much simpler, therefore we will quickly sketch out the procedure.  

Here we use the standard form of the Euclidean $AdS_3$ metric,
\begin{equation}
\mathrm{d}s^2_{AdS} = \frac{\mathcal{R}^2_{AdS}}{z^2} \big(\mathrm{d} t^2 + \mathrm{d} x^2 + \mathrm{d}z^2\big),
\end{equation}
and the   field
\begin{equation}
B = - q\mathcal{R}^2_{AdS}\frac{1}{z^2}\,\mathrm{d}t \wedge \mathrm{d}x.
\end{equation}
The ansatz will read
\begin{equation}  
t(\tau,\sigma) = \tau, \quad x(\tau,\sigma) = x(\sigma), \quad z(\tau,\sigma) = z(\sigma),
\end{equation}
alongside with the boundary condition 
\begin{equation} 
z(\tau,0) = 0.
\end{equation}
Using this ansatz, we find the straight-line action to take the form
\begin{equation}
\begin{split}
S &= \frac{T_0}{2} \int  \mathrm{d}^2\sigma\, \frac{1}{z^2}\Big(1 + x'^2 + z'^2 - 2qx'\Big),
\end{split}
\end{equation}
where $x'\equiv \mathrm{d}x/\mathrm{d}\sigma$ and $z'\equiv \mathrm{d}z/\mathrm{d}\sigma$. The first integral of motion is
\begin{equation}
p = z^{-2} (x'-q),
\end{equation}
supplied with the conformal gauge condition requiring 
\begin{equation}
x'^2 + z'^2 - 1 = 0.
\end{equation}

We again set $p=0$ and find the trivial solution to be 
\begin{equation}
t = \tau,\quad x = q\sigma,\quad z = \sqrt{1-q^2}\,\sigma.
\end{equation}
The pull-back of the target metric will be
\begin{equation}
\mathrm{d}s^2 = R^2 \, \frac{\mathrm{d}\tau^2 + \mathrm{d}\sigma^2}{\sigma^2},
\end{equation}
which is again that of $AdS_2$ with its radius scaled by $R$ in the presence of $B$-field.

Using the solution, the action reads
\begin{equation}
\begin{split}
S &= T_0 \int  \mathrm{d}^2\sigma\, \frac{1}{z^2}\Big(1-q^2\Big) = T_0 \int  \mathrm{d}^2\sigma\, \frac{1}{\sigma^2},
\end{split}
\end{equation}
which vanishes after we remove the divergent term; 
 hence in this case $\Gamma_0 = 0$.

\newpage
\section{Quantum corrections}
In this section we will calculate the one-loop quantum corrections to the minimal surface solutions that we have found in the previous section. The calculation is again quite similar between the two cases and actually produces the same result this time. Thus we will only focus on the more complicated case of the circular loop. We first introduce the quadratic bosonic fluctuations and find their action. Subsequently, the same will be done for the fermionic fluctuations. Finally, the two sectors will be put together and the one-loop  string partition function will be evaluated  using the similarity  with earlier discussed cases.

\subsection{Bosonic fluctuations}
Instead of the Polyakov formalism, which we used in the previous section, we employ the Nambu-Goto formalism to calculate the bosonic fluctuations. We are allowed to do so as the two procedures should produce the same results at  the semiclassical level (see, e.g., \cite{FRADKIN1982413}). The classical action, consisting of the Nambu-Goto and the Wess-Zumino terms, will read 
\begin{equation}
\label{S}
S = T_0 \int \mathrm{d}^{2}\sigma \Big(\sqrt{g} + 
\frac{1}{2}\epsilon^{ij}B_{\mu\nu}\partial_i X^{\mu}\partial_j X^{\nu}\Big),
\end{equation}
where we have again absorbed $\mathcal{R}_{AdS}^2$ into $T_0$, and with the induced metric being
\begin{equation}
g_{ij} = G_{\mu \nu} \frac{\partial X^{\mu}}{\partial \sigma^i}  \frac{\partial X^{\nu}}{\partial \sigma^j} = G_{\mu \nu} \partial_i X^{\mu} \partial_j X^{\nu},
\end{equation}
where the two-dimensional worldsheet is now parameterised by $\sigma^i=(\phi,r)$. Furthermore, if we define the ‘reduced’ coordinates
\begin{equation}
\Bar{r} \equiv R^{-1}r,\quad \Bar{z} \equiv R^{-1}z, \quad \Bar{w} \equiv \sqrt{1-\Bar{r}^2},
\end{equation}
we can write the classical solution as
\begin{equation}
\Bar{z} = \Bar{w} + q, \quad \frac{d \Bar{z}}{dr} = -\frac{1}{R}\frac{\Bar{r}}{\Bar{w}}. 
\end{equation}
The induced metric tensor and the square-root of its determinant will be, respectively,
\begin{equation}
\label{gbar}
\Bar{g}_{ij} = \begin{pmatrix}
  \frac{r^2}{z^2} & 0\\ 
  0 & \frac{1}{\Bar{w}^2 z^2}  
\end{pmatrix}, \quad \sqrt{\bar{g}} = \frac{r}{\Bar{w}z^2}.
\end{equation}

The Nambu-Goto part of the bosonic action reads
\begin{equation}
S^{(0)}_{\mathrm{NG}} = T_0 \int \mathrm{d}^{2}\sigma \sqrt{|\Bar{g}|},
\end{equation}
where $\Bar{g}_{ij}$ is the induced metric of the classical solution defined in (\ref{gbar}). To expand the Nambu-type action to second order in fluctuations around the classical background, $\delta X^{\mu}$, while ensuring a manifestly covariant expression, we will introduce a local power series expansion of $\delta X^{\mu}$ in spacetime vectors, $\xi^\mu$, which are tangent to the spacetime geodesic connecting $\Bar{X}^{\mu}$ and $ \Bar{X}^{\mu} + \delta X^{\mu}$ \cite{Forini_2017}. We then obtain the following expression for the fluctuations 
\begin{equation}
\delta X^{\mu}  = \xi^\mu - \frac{1}{2} \Gamma^\mu_{\rho \nu} \xi^\rho \xi^\nu + \mathcal{O}(\xi^3).
\end{equation}
The perturbed induced metric which also contains the fluctuations around the classical solution is
\begin{equation}
g_{ij} = G_{\mu \nu}(\Bar{X}+\delta X)\,\partial_i(\Bar{X}+\delta X)^{\mu}\partial_j(\Bar{X}+\delta X)^{\nu},
\end{equation}
and thus the perturbation of the metric to second order in $\xi^\mu$ will be
\begin{equation}
\delta g_{ij} = g_{ij} - \Bar{g}_{ij} = \delta_1 g_{ij} + \delta_2 g_{ij},
\end{equation}
where $\delta_1 g_{ij}$ and $\delta_2 g_{ij}$ are the first and second order perturbations respectively. Before we find these perturbations, however, we first introduce the target space vielbeins satisfying $G_{\mu\nu} = E^a_\mu E^a_\nu + E^p_\mu E^p_\nu$, for the $AdS$ index $a=0,1,2$ and the $S^3\times T^4$ index $p=3,\dots,9$. The vielbeins will allow us to define the target space tangent coordinates $\zeta^a = E^a_\mu \xi^{\mu}$ and $\zeta^p = E^p_\mu \xi^{\mu}$, normalised as
\begin{equation}
||\zeta^a||^2 = \int \mathrm{d}^2\sigma \sqrt{\Bar{g}}\, \zeta^a \zeta^a,
\end{equation}
and similarly for $\zeta^p$, and thus produce canonically normalised kinetic terms. Following \cite{Hou_2009}, one finds
\begin{equation}
\begin{split}
\delta_1 g_{ij} &= 2E_{\mu}^a \partial_{(i}\Bar{X}^{\mu} D_{j)} \zeta^{a},\\
\delta_2 g_{ij} &=  D_{i}\zeta^{a} D_{j}\zeta^{a} - R_{acbd} \partial_i \Bar{X}^\mu \partial_j\Bar{X}^\nu E^c_\mu E^d_\nu \zeta^{a} \zeta^{b} + \partial_i \zeta^{p}\partial_j \zeta^{p},
\end{split}
\end{equation}
where the covariant derivative is defined as
\begin{equation}
D_i \zeta^{a} = \partial_i \zeta^{a} + \omega^{ab}_{i} \zeta^{b},\quad \omega^{ab}_{i} = \partial_i \Bar{X}^{\mu} \Omega^{ab}_\mu,
\end{equation}
where $\Omega^{ab}_{\mu}(\Bar{X})$ is the target space spin connection, specifically for $AdS_3$,
\begin{equation}
\Omega^{m2}_{\mu} = - \frac{1}{z} \delta^{m}_{\mu},
\end{equation}
with $m=0,1$ and 
\begin{equation}
R_{acbd} = - \delta_{ab}\delta_{cd} + \delta_{ad}\delta_{bc}.
\end{equation}

It follows then, to quadratic order in $\zeta^{\mu}$, that the Nambu-Goto action expands to
\begin{equation}
S_{\mathrm{NG}} = S^{(0)}_{\mathrm{NG}} + S^{(2)}_{\mathrm{NG}} = S^{(0)}_{\mathrm{NG}} + T_0 (I-J),
\end{equation}
with
\begin{equation}
\label{IJ}
\begin{split}
I &= \frac{1}{2} \int \mathrm{d}^{2}\sigma \sqrt{\Bar{g}} \Bar{g}^{ij} \delta_2 g_{ij},\\
J &=  \frac{1}{8} \int \mathrm{d}^{2}\sigma \sqrt{\Bar{g}} \Big(2\Bar{g}^{ik} \Bar{g}^{jl} - \Bar{g}^{ij} \Bar{g}^{kl}\Big) \delta_1 g_{ij} \delta_1 g_{kl},
\end{split}
\end{equation}
where it has been assumed that every metric present is symmetrical, which is the case in the our example of the worldsheet ending on a circular loop.

Before calculating $I$, we define the mass matrix
\begin{equation}
X_{ab}  = - \Bar{g}^{ij} R_{acbd} \partial_i \Bar{X}^\mu \partial_j\Bar{X}^\nu E^c_\mu E^d_\nu,
\end{equation}
which allows to rewrite $I$ as
\begin{equation}
I = \frac{1}{2} \int \mathrm{d}^{2}\sigma \sqrt{\Bar{g}} (\Bar{g}^{ij}D_{i}\zeta^{a} D_{j}\zeta^{a} + X_{ab}\zeta^a \zeta^b + \Bar{g}^{ij} \partial_i \zeta^{p}\partial_j \zeta^{p}).
\end{equation}
Using the classical solution, one finds that the only non-zero entries of $X_{ab}$ are
\begin{equation}
X_{00} = 1, \quad X_{11} = 1 + \Bar{r}^2, \quad X_{22} = 1 + \Bar{w}^2, \,\,\,\,\,\, X_{12} = X_{21} = \Bar{r} \Bar{w}.   
\end{equation}
The non-zero components of the spin connection in the target space are
\begin{equation}
\Omega^{01}_0 = 1, \quad \Omega^{02}_0 = - \frac{r}{z}, \quad \Omega^{12}_1 = - \frac{1}{z},     
\end{equation}
and thus the only non-trivial covariant derivatives, i.e. not $D_i = \partial_i$, are  
\begin{equation}
\begin{split}
D_0 \zeta^0 = \partial_0 \zeta^0 + \zeta^1 - &\frac{r}{z} \zeta^2, \quad  D_0 \zeta^1 = \partial_0 \zeta^1 - \zeta^0, \quad D_0 \zeta^2 = \partial_0 \zeta^2 + \frac{r}{z} \zeta^0,\\
D_1 \zeta^1 &= \partial_1 \zeta^1  - \frac{1}{z} \zeta^2, \quad D_1 \zeta^2 = \partial_1 \zeta^2  + \frac{1}{z} \zeta^1.
\end{split}
\end{equation}
However, if we now rotate the fields (similar to \cite{Drukker_2000}) such that
\begin{equation}
\label{alpha}
\begin{pmatrix}
\Tilde{\zeta}^1 \\
\Tilde{\zeta}^2 
\end{pmatrix} = 
\begin{pmatrix*}[r]
\mathrm{cos}\,\alpha & -\mathrm{sin}\,\alpha\\
\mathrm{sin}\,\alpha & \mathrm{cos}\,\alpha
\end{pmatrix*}
\begin{pmatrix}
\zeta^1 \\
\zeta^2 
\end{pmatrix}, \quad \mathrm{cos}\,\alpha = \Bar{w}, \quad \mathrm{sin}\,\alpha = \Bar{r}, \quad \frac{\mathrm{d}\alpha}{\mathrm{d}\Bar{r}} = \frac{1}{\Bar{w}},
\end{equation}
the covariant derivatives simplify significantly, as well as $\Tilde{X}_{ab} = \mathrm{diag}(1,1,2)$. Furthermore, in the static gauge, we can fix $\zeta^0 = 0$ and $\Tilde{\zeta}^1 = 0$, as these two longitudinal fluctuations generate a worldsheet diffeomorphism and are therefore zero modes of the fluctuation action. To simplify the expression, we denote the only remaining non-zero, $AdS$, massive fluctuation as $\zeta \equiv \Tilde{\zeta^2}$, we then obtain
\begin{equation}
I = \frac{1}{2} \int \mathrm{d}^{2}\sigma \sqrt{\Bar{g}} \big[\Bar{g}^{ij}\partial_{i}\zeta \partial_{j}\zeta + 2(1+q^2)(\zeta)^2 + \Bar{g}^{ij} \partial_i \zeta^{p}\partial_j \zeta^{p}\big].
\end{equation}

Next, one finds that the only non-vanishing entries of the first order metric perturbation, $\delta_1 g_{ij}$, are
\begin{equation}
\delta_1 g_{00} = q \,\Bar{g}_{00}, \quad \delta_1 g_{11} = q \,\Bar{g}_{11},
\end{equation}
which means $J$ can be easily found to be
\begin{equation}
J = 0.
\end{equation}

Now, putting $I$ and $J$ together and rescaling the quantum fields to absorb the $T_0$ prefactor, we find the second order expansion of the Nambu-type action as
\begin{equation}
\label{ng2}
S_{\mathrm{NG}}^{(2)} = \frac{1}{2} \int \mathrm{d}^{2}\sigma \sqrt{\Bar{g}} \big[\Bar{g}^{ij}\partial_{i}\zeta \partial_{j}\zeta + 2(1+q^2)(\zeta)^2 + \Bar{g}^{ij} \partial_i \zeta^{p}\partial_j \zeta^{p}\big].
\end{equation}

If we now use the same machinery to expand the Wess-Zumino (WZ) term to second order in tangent space fluctuations as in \cite{Callan:1989nz} we find the expression
\begin{equation}
\label{wesszumino}
S_{WZ}^{(2)}=-\frac{T_0}{2} \int \mathrm{d}^2 \sigma\, \epsilon^{ij} H_{\mu \nu \lambda}(\Bar{X}) \partial_i \Bar{X}^{\mu} E^{\nu}_a E^\lambda_b D_j\zeta^a \zeta^b, 
\end{equation}
where $H = \mathrm{d}B$ is the NS-NS 3-form field strength. This is a straight forward calculation, which after imposing the static gauge and rescaling of the quantum fields again, leads to
\begin{equation}
\label{wz2}
S_{WZ}^{(2)}=-\frac{1}{2} \int \mathrm{d}^2 \sigma \sqrt{\Bar{g}}\,\, 4q^2(\zeta)^2. 
\end{equation}

Finally, we find the total second order bosonic action, $S_{B}^{(2)} = S_{NG}^{(2)}+S_{WZ}^{(2)}$, to be 
\begin{equation}
\label{S2B}
S^{(2)}_B = \frac{1}{2} \int \mathrm{d}^{2}\sigma \sqrt{\Bar{g}} \big[\Bar{g}^{ij}\partial_{i}\zeta \partial_{j}\zeta + 2(1-q^2)(\zeta)^2 + \Bar{g}^{ij} \partial_i \zeta^{p}\partial_j \zeta^{p}\big].
\end{equation}
If this result is compared with the one obtained for the case of vanishing $B$-field in \cite{Drukker_2000}, we see the additional $B$-field term in the bosonic action (\ref{S}) only acts to rescale the mass-term of the single massive fluctuation mode. Even more interestingly, however, the scaling factor in this case is $R^{-2} = 1-q^2$, with $R$ being the (dimensionless) radius of the induced $AdS_2$ metric, which can be written as
\begin{equation} \label{eq1} 
\mathrm{d}s^2 = R^2\bigg(\frac{y^2\, \mathrm{d}\phi^2}{1-y^2} +\frac{\mathrm{d}y^2}{(1-y^2)^2} \bigg),
\end{equation}
after we have performed a transformation of the classical solution
\begin{equation} \label{eq} 
\begin{split}
r &= R \frac{y}{1-q\sqrt{1-y^2}},\\
z &= R \frac{\sqrt{(1-q^2) (1-y^2)}}{1-q\sqrt{1-y^2}},
\end{split}
\end{equation}
with $y \in [0,1]$.\footnote{Here, in comparison to section 5 of \cite{Drukker_2000}, $y$ corresponds to $r$ and thus $\sqrt{1-y^2}$ to $w$.} To take this coincidence even further, we now define a new setting; a worldsheet surface ending on a circular loop without the $B$-field but with $AdS$ radius now being $R$, instead of the unit radius used before. Again, this is the case already studied in \cite{Drukker_2000} and as such it is easy to show that $R$ indeed only changes the unit radius, second order bosonic action by rescaling the mass-matrix $X \to X/R^2$, and thus resulting in the same action as found here (\ref{S2B}).

This   implies   that   the  bosonic fluctuation part  of the 
one-loop  correction to the string partition function corresponding to a minimal surface ending on a unit radius circular loop for $q\not=0$ (i.e.
 in the presence of NS-NS flux)  can be found  by  calculating the correction for a case of the surface ending on a circular loop of radius $R$ for $q=0$ (i.e. pure R-R flux). We will now go on to show that this applies to the  to fermionic case  as well.

\subsection{Fermionic fluctuations}
Following \cite{wulff2013type,Cveti__2000}, the quadratic term in the fermionic part of the Green-Schwarz action (using Minkowski signature) is
\begin{equation}
\label{Faction}
S_{F}^{(2)} = T_0 \int \mathrm{d}^2 \sigma\, \mathcal{L}_F  = -iT_0 \int \mathrm{d}^2 \sigma \big(\sqrt{-\bar{g}}\bar{g}^{ij} \delta^{IJ} + \epsilon^{ij}\sigma_3^{IJ}\big)\Bar{\theta}^I \rho_i D_j \theta^J,
\end{equation}
where $\theta^I$ are two left 10d Majorana-Weyl spinors, $\sigma_i^{IJ}$ are Pauli matrices, $\rho_i = \partial_i \Bar{X}^\mu E^a_\mu \Gamma_a$ are the 10d Dirac matrices pulled back onto the worldsheet, and the covariant derivatives take the form
\begin{equation}
\label{fermionic derivatives}
D_i \theta^I = \Big[\Big(\partial_i + \frac{1}{4}\omega^{ab}_i \Gamma_{ab}\Big) \delta^{IJ} +\frac{1}{8}e^{a}_i H_{abc} \Gamma^{bc}\sigma_3^{IJ} +  \frac{1}{8\times 3!}F_{abc}\Gamma^{abc}e^{d}_i\Gamma_d \sigma_1^{IJ} \Big]\theta^J,
\end{equation}
with $e^a_i = \partial_i \Bar{X}^\mu E^a_\mu$ and the the NS-NS and R-R 3-form fluxes, respectively, being
\begin{equation} \label{fluxes}
\begin{split}
H &= 2q\, (\omega_{AdS_3} + \omega_{S^3}),\\
F &= 2\Bar{q}\, (\omega_{AdS_3} + \omega_{S^3}),
\end{split}
\end{equation}
where $\omega_{AdS_3}$ and $\omega_{S^3}$ are volume forms on $AdS_3$ and $S^3$ of unit radius, with $\Bar{q}=\sqrt{1-q^2}$, and the square of the common radius of both spaces being absorbed by the string tension again. The $S^3$ subspace does not contribute to the $H$-term in the covariant derivative, however, it does contribute to the $F$-term.

One then finds the covariant derivatives to be
\begin{equation}
\begin{split}
D_0  &= \partial_0 + \frac{1}{2}\Gamma_{0}\bigg(\Gamma_{1}-\frac{r}{z}\Gamma_{2}\bigg) + \frac{r}{2z}\Gamma_{12}\big(\Bar{q}P\sigma_1 + q\sigma_3\big),  \\
D_1  &= \partial_1 - \frac{1}{2z}\Gamma_{12} +  \frac{1}{2z}\Gamma_0\bigg( \Gamma_2 + \frac{\Bar{r}}{\Bar{w}} \Gamma_1\bigg)\big(\Bar{q}P\sigma_1 + q\sigma_3\big),\\
\end{split}
\end{equation}
where the spinor indices are implicit and 
\begin{equation}
P=\frac{1+\Gamma_{012345}}{2}
\end{equation}
is a projection operator. To simplify these expressions, we will use the following rotation matrix as in \cite{Drukker_2000}
\begin{equation}
S = \mathrm{exp}\Big(\frac{\alpha}{2}\,\Gamma_{12}\Big),
\end{equation}
with $\alpha$ being defined in (\ref{alpha}). We may then rewrite the 2d pulled back matrices using $S$ as
\begin{equation}
\begin{split}
\rho_0  &=  \frac{r}{z} \Gamma_0 = e^{\alpha}_0 S\Gamma_\alpha S^{-1},\\
\rho_1  &=  \frac{1}{z} \Gamma_1 - \frac{\Bar{r}}{z\Bar{w}}\Gamma_2 = e^{\alpha}_1 S\Gamma_\alpha S^{-1},
\end{split}
\end{equation}
with the index $\alpha = 0,1$, and $e^\alpha_i$ being the zweibeins of the induced metric. Furthermore, the covariant derivatives can be rewritten using $S$ as
\begin{equation}
\begin{split}
\label{devs}
D_0 &= S\Big[\hat{\nabla}_0 + \frac{qr}{2z}\Gamma_{02} + \frac{r}{2z}\Gamma_{12}\big(\Bar{q}P\sigma_1 + q\sigma_3\big)\Big]S^{-1},  \\
D_1 &= S\Big[\hat{\nabla}_1 + \frac{q}{2z\Bar{w}}\Gamma_{12} + \frac{1}{2z\Bar{w}}\Gamma_{02}\big(\Bar{q}P\sigma_1 + q\sigma_3\big)\Big]S^{-1},  \\
\end{split}
\end{equation}
where $\hat{\nabla}_i$ is the covariant derivative with spinor worldsheet connection\footnote{Explicitly, $\hat{\nabla}_i = \partial_i + \frac{1}{4} \Omega_i^{\alpha \beta} \Gamma_{\alpha \beta}$ with $\Omega_i^{\alpha \beta} = e^{\alpha j} \partial_{[i}e^\beta_{j]} - e^{\beta j} \partial_{[i}e^\alpha_{j]} - e^{\alpha j} e^{\beta k} e^{\gamma}_i \partial_{[j} e^{\gamma}_{k]} $ and $\alpha, \beta \in \{0,1\}.$},
\begin{equation}
\label{worldsheet derivative}
\hat{\nabla}_0 = \partial_0 + \frac{1+q\Bar{w}}{2\Bar{z}}\Gamma_{01}, \quad \hat{\nabla}_1 = \partial_1.
\end{equation}

If one now defines $\mathbf{\Psi}$ such that $\boldsymbol{\theta} = S \mathbf{\Psi}$ and chooses the kappa gauge $\Psi^1 = \Psi^2 \equiv \Psi$, one finds that the second order fermionic action (\ref{Faction}) simplifies significantly as the second and fourth terms in the square brackets of (\ref{devs}) exactly cancel after summation, thus leaving one with
\begin{equation}
\label{fermionic lagrangian}
\mathcal{L}_F = -2i\sqrt{-\bar{g}}\,\Bar{\Psi} \big(\bar{g}^{ij}e^\alpha_i \Gamma_\alpha \hat{\nabla}_j - \Bar{q}\Gamma_{012}P \big) \Psi.
\end{equation}
If we now diagonalise $P$, we will be able to rewrite the equation using two sixteen-component Majorana-Weyl spinors, $\psi^1$ and $\psi^2$, whose eigenvalues with
respect to the projector $P$ are one and zero, respectively. Therefore we obtain
\begin{equation}
\begin{split}
\mathcal{L}_{F,1} &= -2i\sqrt{-\bar{g}}\,\Bar{\psi}^1 \big(\bar{g}^{ij}e^\alpha_i \Gamma_\alpha \hat{\nabla}_j - \Bar{q}\Gamma_{012} \big) \psi^1, \\
\mathcal{L}_{F,2} &= -2i\sqrt{-\bar{g}}\,\Bar{\psi}^2 \bar{g}^{ij}e^\alpha_i \Gamma_\alpha \hat{\nabla}_j \psi^2.
\end{split}
\end{equation}

Again, interestingly, we observe a rescaling of the (linear) mass term by $R^{-1} = \bar{q}$, in the comparison to the pure R-R flux case, and that one can find the exact same result by working with a background metric whose radius is $R$ instead of $1$, and with vanishing $H$-field. This happens as the two crossterms that cancel against each other, which arise in our previous case, both depend on $q$ which we set to zero now (except for in $\Bar{q}$ or $R$). It is also important, that the $F$-field stays the same as in the previous calculation above (in tangent coordinates) since now
\begin{equation}
F = 2R^2 (\omega_{AdS_3}+ \omega_{S^3}) = 2\Bar{q} \Big(\omega_{AdS_3}^{(R)} + \omega_{S^3}^{(R)}\Big),
\end{equation}
where $\omega_{AdS_3}^{(R)}$ and $\omega_{S^3}^{(R)}$  are the volume forms on $AdS_3$ and $S^3$ of radius $R$, respectively.

\subsection{One-loop partition function}
Given that we have shown that both the bosonic and fermionic second order actions are identical in the case of the metric having unit radius and $B$-field being present, and in the case of the radius being $R=(1-q^2)^{-\frac{1}{2}}$ with no $B$-field, we should be able to use the latter to evaluate the former. This has been explained in \cite{Giombi_2020}, in relation to discussing how a change in the radius of the metric impacts the semiclassical partition function. It was argued that if 
\begin{equation}
\Gamma_1^{(0)} = -\frac{1}{2}\,\mathrm{log}\,2\pi - \mathrm{log} \sqrt{\frac{T_0}{2\pi g_s^2}} = - \mathrm{log} \frac{\sqrt{T_0}}{g_s}
\end{equation}
is the unit-radius one-loop correction, $g_s$ is the string coupling, and $T_0$ is again defined to have absorbed the square of the $AdS$ radius, then one should in our case find the scaled correction to be 
\begin{equation}
\Gamma_1 = \Gamma_1^{(0)} - \mathrm{log}\,R = -\mathrm{log} \frac{\sqrt{T_0 R^2}}{g_s}.
\end{equation}
Therefore, by assuming the mentioned $B/R$ correspondence of the two cases (at least to one-loop order), we should find the desired one-loop correction of the non-zero $B$-field case to read
\begin{equation}
\Gamma_1 = -\mathrm{log} \frac{\sqrt{T}}{g_s},
\end{equation}
where we have defined the scaled tension 
\begin{equation}
T \equiv T_0 R^2 = \frac{T_0}{1-q^2}.
\end{equation}
As discussed previously, this result applies to both the single circle and straight line Wilson loops. Using $T$, we can also express the classical action of the single circle as $\Gamma_0 = -2\pi T$.

Finally, we obtain the one-loop semiclassical partition function, $W_1$, for the unit-radius circular Wilson loop in the presence of $B$-field as 
\begin{equation}
 W_1  = \frac{\sqrt{T}}{g_s}e^{2\pi T},
\end{equation}
and for the straight line case, with $\Gamma_0 = 0$, simply as
\begin{equation}
 W_1  = \frac{\sqrt{T}}{g_s}.
\end{equation}

\section{Conclusion}
In conclusion, we have revisited the problem of calculating the semiclassical partition function for circular  and straight-line Wilson  loops  in  IIB string on an $AdS_3 \times S^3 \times T^4$ background in the presence of both the R-R and NS-NS 3-form fluxes. We have shown  that there exists a single circle solution even in the presence of the NS-NS   $B$-field, which was  mistaken for a case of ending on two concentric circles in the  previous work  \cite{Hernandez_2020}. For this new  classical solution, which  describes an  $AdS_2$ minimal surface, we found   the quadratic fluctuation Lagrangian for the bosonic and fermionic  string coordinates. Unlike what was claimed in \cite{Hernandez_2020}, we found that that  the dependence on the NS-NS flux   parameter  contribution to the semiclassical partition function does not  drop out  but instead acts to rescale the masses of both the bosonic and fermionic fluctuations. 
 
Thus at both  the classical and one-loop level, we have observed that the dependence on the NS-NS flux parameter $q$ can be captured  by  simply rescaling  the  $AdS$ radius by a factor $R =(1-q^2)^{-\frac{1}{2}}$. It would be interesting to see if similar rescaling works in the case of other Wilson loops like, e.g., anti-parallel  lines.

\section*{Acknowledgements}
I would like to thank  Arkady  Tseytlin for taking me on to do a summer research project under his supervision, which later morphed into my MSci project, resulting in this work. I am very grateful for the interesting topic I had been given, as well as for all the challenging discussions we have had throughout the year and all the help he has kindly provided.

\appendix
\section{Classical action} \label{appendix1}
If we were to invert the coordinate transformation defined in (\ref{eqcoordinates}) which is
\begin{equation} 
\begin{split}
r = \frac{e^v}{\sqrt{1+u^2}}, \quad z = \frac{ue^v}{\sqrt{1+u^2}} + qR,
\end{split}
\end{equation}
one would find
\begin{equation}
u = \frac{z-qR}{r}, \quad v = \mathrm{log}\sqrt{r^2 + (z-qR)^2},
\end{equation}
which leads to the following form of the $AdS$ metric
\begin{equation}
\mathrm{d}s^2_{AdS} = \frac{\mathcal{R}_{AdS}^2}{\big(u + qRe^{-v}\sqrt{1+u^2}\big)^2} \Big[\mathrm{d}\phi^2 + \frac{\mathrm{d} u^2}{1+u^2} + (1+u^2)\mathrm{d}v^2 \Big].
\end{equation}

These coordinates are useful since
\begin{equation}
\mathrm{d}v = \frac{\mathrm{d}r + u \mathrm{d}z}{r(1+u^2)},
\end{equation}
which lets us naturally express $B$-field defined in (\ref{bfield}) as
\begin{equation}
\begin{split}
B &= q \mathcal{R}_{AdS}^2 \frac{r}{z^2}\big[\mathrm{d}r + u \mathrm{d}z\big] \wedge \mathrm{d}\phi\\
&= q \mathcal{R}_{AdS}^2 \frac{r^2}{z^2}(1+u^2)\mathrm{d}v \wedge \mathrm{d}\phi\\
&= -q \mathcal{R}_{AdS}^2 \frac{1+u^2}{\big(u + qRe^{-v}\sqrt{1+u^2}\big)^2}\mathrm{d}\phi\wedge \mathrm{d}v,
\end{split}
\end{equation}
which satisfies 
\begin{equation}
\begin{split}
H &= dB\\
&=  \frac{2 q \mathcal{R}_{AdS}^2}{\big(u + qRe^{-v}\sqrt{1+u^2}\big)^3}\,\mathrm{d}\phi\wedge \mathrm{d}v \wedge \mathrm{d}u\\
&= 2 q \mathcal{R}_{AdS}^2 \,\omega_{AdS},
\end{split}
\end{equation}
where $\omega_{AdS}$ is again the volume form on $AdS_3$ of unit radius.

Now in these new coordinates we can use the following ansatz which is consistent with the equations of motion, 
\begin{equation}
\phi = \tau, \quad u = u(\sigma), \quad v = v(\sigma),
\end{equation}
and the Lagrangian becomes
\begin{equation}
\begin{split}
S &= \frac{T_0}{2} \int  \mathrm{d}^2\sigma\, \frac{1}{\big(u + qRe^{-v}\sqrt{1+u^2}\big)^2} \Big[1 + \frac{u'^2}{1+u^2} + (1+u^2)v'^2 - 2q(1+u^2)v' \Big].
\end{split}
\end{equation}
The conformal gauge condition reads
\begin{equation}
\frac{u'^2}{1+u^2} + (1+u^2)v'^2 - 1 = 0,
\end{equation}
and the equation of motion for $v$ is
\begin{equation}
\label{veq}
\frac{d}{d\sigma}\Bigg[\frac{2(1+u^2)(v'-q)}{\Big(u + qRe^{-v}\sqrt{1+u^2}\Big)^2}\Bigg] -  \frac{2qRe^{-v}\sqrt{1+u^2}}{\Big(u + qRe^{-v}\sqrt{1+u^2}\Big)^3} \Bigg[1 + \frac{u'^2}{1+u^2} + (1+u^2)(v'^2 - 2qv') \Bigg] = 0.
\end{equation}
These two equations are again solved by
\begin{equation}
v = \mathrm{log}\,R, \quad u = \mathrm{sinh}(\sigma - \mathrm{arctanh}\,q),
\end{equation}
with $\sigma \in [0,\infty)$, as then $v' = 0$ implies
\begin{equation}
u'^2 = 1 + u^2,
\end{equation}
which means $u = \mathrm{sinh}(\sigma + C)$, where $C$ is the integration constant. After imposing boundary conditions $z(\sigma=0) = 0$ implying that 
\begin{equation}
\frac{u(0)}{\sqrt{1+u^2(0)}} + q = 0,
\end{equation}
since (after solving the equations)
\begin{equation}
z = \frac{Ru}{\sqrt{1+u^2}} + qR.
\end{equation}
Then this is solved by taking $u(0) = -q/\sqrt{1-q^2}$ or by taking $C = - \mathrm{arctanh}\,q$. This also solves the equation for $v$ as (\ref{veq}) becomes
\begin{equation}
\label{veqq}
\frac{d}{d\sigma}\Bigg[\frac{1+u^2}{\Big(u + q\sqrt{1+u^2}\Big)^2}\Bigg] +   \frac{2\sqrt{1+u^2}}{\Big(u + q\sqrt{1+u^2}\Big)^3} = 0.
\end{equation}
Now if one notices that 
\begin{equation}
\begin{split}
u + q\sqrt{1+u^2} &= \mathrm{sinh}(\sigma- \mathrm{arctanh}\,q) + q \mathrm{cosh}(\sigma - \mathrm{arctanh}\,q)\\  
&= R^{-1} \mathrm{sinh}\,\sigma,
\end{split}
\end{equation}
they find
\begin{equation}
\frac{1+u^2}{\Big(u + q\sqrt{1+u^2}\Big)^2} = \frac{R^2\mathrm{cosh}^2(\sigma- \mathrm{arctanh}\,q)}{\mathrm{sinh}^2\,\sigma},
\end{equation}
so then 
\begin{equation}
\frac{d}{d\sigma} \Bigg[\frac{R^2\mathrm{cosh}^2(\sigma- \mathrm{arctanh}\,q)}{\mathrm{sinh}^2\,\sigma}\Bigg] = - \frac{2R^3 \mathrm{cosh}(\sigma- \mathrm{arctanh}\,q)}{\mathrm{sinh}^3\,\sigma},
\end{equation}
which cancels against the second term in (\ref{veqq}) which is
\begin{equation}
\frac{2\sqrt{1+u^2}}{\Big(u + q\sqrt{1+u^2}\Big)^3} = \frac{2R^3 \mathrm{cosh}(\sigma- \mathrm{arctanh}\,q)}{\mathrm{sinh}^3\,\sigma},
\end{equation}
hence the solution also satisfies the equations of motion of $v$.

Hence the final form of the action is
\begin{equation}
\begin{split}
S &= T_0  \int  \mathrm{d}^2\sigma\, \frac{1}{\big(u + q\sqrt{1+u^2}\big)^2}\\
&= T_0 R^2 \int  \mathrm{d}^2\sigma\, \frac{1}{\mathrm{sinh}^2(\sigma)}\\
&= -2\pi T_0 R^2,
\end{split}
\end{equation}
as shown previously.

\section{Bosonic fluctuations}
If we invert the relationship defining the rotation of the bosonic fluctuations defined in (\ref{alpha}) and subsequently apply the static gauge condition (i.e. $\zeta^0 = 0$, $\Tilde{\zeta}^1 = 0$, and $\Tilde{\zeta}^2 = \zeta$), then we may write 
\begin{equation}
\begin{split}
\zeta^1 &= \bar{w} \Tilde{\zeta}^1 + \bar{r} \Tilde{\zeta}^2 \, \to \, \bar{r} {\zeta},\\
\zeta^2 &= \bar{w} \Tilde{\zeta}^2 - \bar{r} \Tilde{\zeta}^1 \, \to \, \bar{w} {\zeta}.
\end{split}
\end{equation}
This allows us to explicitly find the non-trivial bosonic covariant derivatives
\begin{equation}
\begin{split}
D_0 \zeta^0 = \frac{qr}{z} {\zeta}, \quad  D_0 \zeta^1 = \bar{r}\,\partial_0 {\zeta}, \quad D_0 \zeta^2 = \bar{w}\, \partial_0 {\zeta},\\
D_1 \zeta^1 = \bar{r} \, \partial_1 \zeta  + \frac{q}{z} \zeta, \quad D_1 \zeta^2 = \bar{w} \, \partial_1 \zeta  - \frac{q\bar{r}}{z\bar{w}} \zeta,
\end{split}
\end{equation}
and thus the non-trivial kinetic terms of the bosonic action
\begin{equation}
\begin{split}
\Bar{g}^{00} \Big[\big(D_0 \zeta^0\big)^2 + \big(D_0 \zeta^1\big)^2 + \big(D_0 \zeta^2\big)^2\Big] &= \Bar{g}^{00} \big(\partial_0 \zeta\big)^2 + q^2 (\zeta)^2,\\
\Bar{g}^{11} \Big[\big(D_1 \zeta^1\big)^2  + \big(D_1 \zeta^2\big)^2\Big] &= \Bar{g}^{11} \big(\partial_1 \zeta\big)^2 + q^2 (\zeta)^2,
\end{split}
\end{equation}
resulting in the Nambu-Goto part of the second order bosonic action in (\ref{ng2}).

For the second order Wess-Zumino term in the bosonic part of the action as in (\ref{wesszumino}), we have  
\begin{equation}
S_{WZ}^{(2)}=-\frac{T_0}{2} \int \mathrm{d}^2 \sigma\, \epsilon^{ij} H_{abc} e^c_i  D_j\zeta^a \zeta^b = -\frac{T_0}{2} \int \mathrm{d}^2 \sigma\,L^{(2)}_{WZ}, 
\end{equation}
with $H_{abc} = 2q\, \epsilon_{abc}$, $e^a_0 = \Big(\frac{r}{z}, 0, 0\Big)$, and $e^a_1 = \Big(0, \frac{1}{z}, -\frac{\bar{r}}{z\bar{w}}\Big)$. Thus the Lagrangian term defined above is
\begin{equation}
\begin{split}
L^{(2)}_{WZ} &= 4q\, \epsilon_{abc}  e^{c}_{[0}  D_{1]}\zeta^a \zeta^b \\
&= 4q \Big(e^0_0   D_{1}\zeta^{[1} \zeta^{2]} - e^1_1
  D_{0}\zeta^{[2} \zeta^{0]} - e^2_1
    D_{0}\zeta^{[0} \zeta^{1]} \Big),
\end{split}
\end{equation}
where
\begin{equation}
\begin{split}
D_{1}\zeta^{[1} \zeta^{2]} &= \frac{q}{2z\bar{w}} (\zeta)^2,\\
D_{0}\zeta^{[2} \zeta^{0]} &= -\frac{qr\Bar{w}}{2z} (\zeta)^2,\\
D_{0}\zeta^{[0} \zeta^{1]} &= \frac{qr\Bar{r}}{2z} (\zeta)^2,
\end{split}
\end{equation}
with $\sqrt{\bar{g}} = \frac{r}{z^2 \Bar{w}}$ again. After substituting in the corresponding values, one finds
\begin{equation}
L^{(2)}_{WZ} = 4q^2 \sqrt{\bar{g}} (\zeta)^2,
\end{equation}
and the second order Wess-Zumino term follows in (\ref{wz2}).

\section{Fermionic fluctuations}
The spinor worldsheet covariant derivatives introduced in (\ref{worldsheet derivative}) utilise the worldsheet zweibeins $e_0^\alpha = \Big(\frac{r}{z},0 \Big)$ and $e_1^\alpha = \Big(0, \frac{1}{z\bar{w}} \Big)$, which also satisfy $\partial_0 e^\alpha_i = 0$. The derivative itself is defined as
\begin{equation}
\begin{split}
\hat{\nabla}_i = \partial_i + \frac{1}{4} \Omega_i^{\alpha \beta} \Gamma_{\alpha \beta},
\end{split}
\end{equation}
where
\begin{equation}
\begin{split}
\Omega_i^{\alpha \beta} &= e^{\alpha j} \partial_{[i}e^\beta_{j]} - e^{\beta j} \partial_{[i}e^\alpha_{j]} - e^{\alpha j} e^{\beta k} e^{\gamma}_i \partial_{[j} e^{\gamma}_{k]}.
\end{split}
\end{equation}
Due to its antisymmetric nature, one only needs to concern themselves with finding $\Omega_0^{01}$ and $\Omega_1^{01}$. Starting with the former,
\begin{equation}
\begin{split}
\Omega_0^{01} &= e^{00} \underbrace{\partial_{[0}e^1_{0]}}_{=0} - e^{11} \underbrace{\partial_{[0}e^0_{1]}}_{= -\frac{1}{2}\partial_1 e^0_0}  - e^{00} e^{11} e^{0}_0 \underbrace{\partial_{[0} e^{0}_{1]}}_{= -\frac{1}{2}\partial_1 e^0_0}\\
&= e^{11}\partial_1 e^0_0 = \frac{1+q\bar{w}}{\bar{z}},
\end{split}
\end{equation}
and subsequently with the latter,
\begin{equation}
\begin{split}
\Omega_1^{01} &= e^{00} \underbrace{\partial_{[1}e^1_{0]}}_{=0} - e^{11} \underbrace{\partial_{[1}e^0_{1]}}_{=0}  - e^{00} e^{11} e^{0}_0 \underbrace{\partial_{[0} e^{0}_{1]}}_{= 0}\\
&= 0,
\end{split}
\end{equation}
as $\partial_0 e^1_1 = 0$ and $e^1_0 = 0$. This in turn leads to the expression (\ref{worldsheet derivative}).

Following the convention of using $\Gamma_{(a} \Gamma_{b)} = \eta_{ab}$ for the 10d gamma matrices with $\eta = (-+\cdots+)$ as in \cite{Drukker_2006}, the 3-form NS-NS terms in the spinor covariant derivative terms of (\ref{fermionic derivatives}) are
\begin{equation}
\begin{split}
\frac{1}{8}e^a_0 H_{abc} \Gamma^{bc} &= \frac{1}{4}e^0_0 H_{012} \Gamma^{12} = \frac{qr}{2z} \Gamma_{12}, \\
\frac{1}{8}e^a_1 H_{abc} \Gamma^{bc} &= \frac{1}{4}H_{012} \big(e^1_1 \Gamma^{20} + e^2_1 \Gamma^{01} \big) = \frac{q}{2z}\Gamma_0 \Big( \Gamma_2 + \frac{\Bar{r}}{\Bar{w}}\Gamma_1  \Big), 
\end{split}
\end{equation}
and the R-R 3-form terms
\begin{equation}
\begin{split}
\frac{1}{8\times 3!}F_{abc}\Gamma^{abc} e^d_0 \Gamma_{d} &=  \frac{\Bar{q}r}{4z} \big(\Gamma^{012} + \Gamma^{345} \big)\Gamma_0\\
&= \frac{\Bar{q}r}{4z} \Gamma_{12} \big(1 + \Gamma_{012345} \big)\\
&= \frac{\Bar{q}r}{2z} \Gamma_{12} P, \\
\frac{1}{8\times 3!}F_{abc}\Gamma^{abc} e^d_1 \Gamma_{d} &= \frac{\bar{q}}{4} \big(\Gamma^{012} + \Gamma^{345} \big) \Big(\frac{1}{z} \Gamma_1 - \frac{\bar{r}}{z\bar{w}}\Gamma_2 \Big) \\
&= - \frac{\bar{q}}{4z} \Big(\Gamma_1 - \frac{\bar{r}}{\bar{w}}\Gamma_2 \Big) \big(\Gamma_{012} + \Gamma_{345} \big) \\
&= - \frac{\bar{q}}{2z} \Big(\Gamma_1 - \frac{\bar{r}}{\bar{w}}\Gamma_2 \Big) \Gamma_{012} P \\
&= \frac{\bar{q}}{2z} \Gamma_{0} \Big(\Gamma_2 + \frac{\bar{r}}{\bar{w}}\Gamma_1 \Big) P.
\end{split}
\end{equation}

Using the expressions defined in (\ref{alpha}), the rotation matrix used to locally Lorentz rotate the GS spinors, i.e. $\boldsymbol{\theta} = S \mathbf{\Psi}$,
to simplify the GS action (\ref{Faction}) into the action of 2d fermions is $S = \mathrm{exp}\Big(\frac{\alpha}{2}\Gamma_{12}\Big)$, for which we have $[S,\Gamma_{12}]= 0$, $[S, \Gamma_a] = 0$ with $a\neq 1,2$, and $\Gamma_{1,2} S = S^{-1} \Gamma_{1,2}$. Investigating the action term by term, we first find that
\begin{equation}
\begin{split}
\frac{1}{2} \Gamma_0 \Big(\Gamma_1 - \frac{r}{z} \Gamma_2 \Big) &= \frac{1}{2} \Gamma_0 \Big(\Gamma_1 - \frac{r}{z} \Gamma_2 \Big) S S^{-1} \\
&= S\, \frac{1}{2} \Gamma_0 S^{-2} \Big(\Gamma_1 - \frac{r}{z} \Gamma_2 \Big)S^{-1}\\
&= S\, \frac{1}{2} \Gamma_0 \Big[\bar{w}\Gamma_1 + \bar{r}\Gamma_2 - \frac{r}{z} \big(\bar{w}\Gamma_2 - \bar{r}\Gamma_1\big) \Big]S^{-1}\\
&= S\, \frac{1}{2\Bar{z}} \Gamma_0 \Big[\big(1+q\bar{w} \big)\Gamma_1 + q\Bar{r}\Gamma_2 \Big]S^{-1}\\
&= S\, \bigg[\frac{1}{4}\Omega_0^{\alpha \beta}\Gamma_{\alpha \beta} + \frac{qr}{2z} \Gamma_{02} \bigg]S^{-1},
\end{split}
\end{equation}
where again $\Omega_i^{\alpha \beta}$ is the worldsheet spin connection. Now since $S = S(r)$ one finds
\begin{equation}
\begin{split}
(\partial_1 S) S^{-1} &= S \frac{1}{2} \Gamma_{12} (\partial_1 \alpha) S^{-1}=  S \frac{1}{2w} \Gamma_{12} S^{-1}, 
\end{split}
\end{equation}
and thus 
\begin{equation}
\partial_1 = S \Big(\partial_1 + \frac{1}{2w} \Gamma_{12}\Big) S^{-1},
\end{equation}
additionally,
\begin{equation}
\begin{split}
\frac{1}{2}\Gamma_0\Big(\Gamma_2 + \frac{\bar{r}}{\bar{w}} \Gamma_1 \Big) &= \frac{1}{2\bar{w}}\Gamma_0 S^{2}\Gamma_2 = \frac{1}{2\bar{w}}\Gamma_0 S\Gamma_2S^{-1} = S\frac{1}{2\bar{w}}\Gamma_{02}S^{-1}.
\end{split}
\end{equation}

Finally, we go over the explicit evaluation of the fermionic Lagrangian. After defining $\mathbf{\Psi}$ such that $\boldsymbol{\theta} = S \mathbf{\Psi}$ and choosing the kappa gauge $\Psi^1 = \Psi^2 \equiv \Psi$, we can rewrite the Lagrangian as
\begin{equation}
\begin{split}
\mathcal{L}_F &= -i\big(\sqrt{-\bar{g}}\bar{g}^{ij} \delta^{IJ} + \epsilon^{ij}\sigma_3^{IJ}\big)\Bar{\theta}^I \rho_i D_j \theta^J\\
&= -i\bar{\Psi} \, \mathrm{Tr}\Big[\mathbb{J}_2\,\big(\sqrt{-\bar{g}}\bar{g}^{ij} \mathbb{1} + \epsilon^{ij}\sigma_3 \big) e^\alpha_i \Gamma_\alpha S^{-1} D_j S  \Big] \Psi,
\end{split}
\end{equation}
where $\mathbb{J}_2^{IJ} = 1$ is an all-ones matrix of size 2 and the trace is taken over the $I$,$J$ indices. Now using the fact that the only non-zero traced-over parts of the action will include the following terms,
\begin{equation}
\mathrm{Tr}\big[\mathbb{J}_2 \big] = 2, \quad \mathrm{Tr}\big[\mathbb{J}_2 \sigma_1 \big] = 2, \quad
\mathrm{Tr}\big[(\sigma_3)^2 \big] = 2,
\end{equation}
one immediately finds
\begin{equation}
\begin{split}
\mathcal{L}_F = -2i \bar{\Psi} &\Big[\sqrt{-\bar{g}}\bar{g}^{00}e^0_0\Gamma_0 \big(\hat{\nabla}_0 + \frac{q}{2}e^0_0 \Gamma_{02} + \frac{\Bar{q}}{2}e^0_0 \Gamma_{12} P \big)\\ &+ \sqrt{-\bar{g}}\bar{g}^{11}e^1_1\Gamma_{1} \big(\hat{\nabla}_1 + \frac{q}{2}e^1_1 \Gamma_{12} + \frac{\Bar{q}}{2}e^1_1 \Gamma_{02}P \big) \\
&+ e^0_0\Gamma_0 \frac{q}{2} e^1_1 \Gamma_{02} - e^1_1\Gamma_1 \frac{q}{2} e^0_0 \Gamma_{12}
\Big] \Psi,
\end{split}
\end{equation}
which after regrouping the derivative terms and remembering that $\sqrt{-\Bar{g}} = e^0_0 e^1_1$ yields
\begin{equation}
\begin{split}
\mathcal{L}_F = -2i \sqrt{-\bar{g}}\, \bar{\Psi} \Big[\bar{g}^{ij}e^\alpha_i\Gamma_\alpha \hat{\nabla}_j &+ \frac{q}{2} \Gamma_{2} - \frac{\Bar{q}}{2} \Gamma_{012} P \\ &+ \frac{q}{2} \Gamma_{2} - \frac{\Bar{q}}{2} \Gamma_{012}P \\
&- \frac{q}{2} \Gamma_{2} -  \frac{q}{2}\Gamma_{2}\Big] \Psi.
\end{split}
\end{equation}
Here we can notice that terms originating from the $H$-contributions to the spinor covariant derivatives in (\ref{fermionic derivatives}) exactly cancel against the residue terms originating from the spin connection contributions. In the end, one is left with the final form of the second order fermionic Lagrangian found in (\ref{fermionic lagrangian}), with only the R-R $F$-terms contributing to the final form of the mass-term.

\newpage

\bibliographystyle{JHEP}
\bibliography{main}

\end{document}